\documentclass[12pt]{article}

\usepackage{graphicx}
\usepackage{amsmath}
\usepackage{amssymb}
\allowdisplaybreaks

\begin{document}

\baselineskip=17.5pt plus 0.2pt minus 0.1pt

\renewcommand{\theequation}{\arabic{equation}}
\renewcommand{\thefootnote}{\fnsymbol{footnote}}
\makeatletter
\def\CR{\nonumber \\}
\def\pt{\partial}
\def\be{\begin{equation}}
\def\ee{\end{equation}}
\def\bea{\begin{eqnarray}}
\def\eea{\end{eqnarray}}
\def\bead{\be\begin{aligned}}
\def\eead{\end{aligned}\ee}
\def\eq#1{(\ref{#1})}
\def\la{\langle}
\def\ra{\rangle}
\def\hyp{\hbox{-}}

\begin{titlepage}
\title{\hfill\parbox{4cm}{ \normalsize YITP-11-51}\\
\vspace{1cm} Tensor models and hierarchy of n-ary algebras}
\author{
Naoki {\sc Sasakura}\thanks{\tt sasakura@yukawa.kyoto-u.ac.jp}
\\[15pt]
{\it Yukawa Institute for Theoretical Physics, Kyoto University,}\\
{\it Kyoto 606-8502, Japan}}
\date{}
\maketitle
\thispagestyle{empty}
\begin{abstract}
\normalsize
Tensor models are generalization of matrix models, and are studied as models of 
quantum gravity.
It is shown that the symmetry of the rank-three tensor models is generated by a hierarchy of 
n-ary algebras starting from the usual commutator, and 
the 3-ary algebra symmetry reported in the previous paper 
is just a single sector of the whole structure. 
The condition for the Leibnitz rules of the n-ary algebras is 
discussed from the perspective of
the invariance of the underlying algebra under the n-ary transformations.
It is shown that the n-ary transformations which keep the underlying algebraic structure
invariant 
form closed finite n-ary Lie subalgebras.  
It is also shown that,
in physical settings, the 3-ary transformation practically generates only local infinitesimal 
symmetry transformations,
and  the other more non-local infinitesimal symmetry transformations of the tensor models
are generated by higher n-ary transformations.
\end{abstract}
\end{titlepage}

\section{Introduction}
\label{sec:intro}
Tensor models have originally been introduced \cite{Ambjorn:1990ge,Sasakura:1990fs,Godfrey:1990dt}
to describe the simplicial quantum gravity in general dimensions higher than two. 
The formulation has later been extended to describe spin foam and loop quantum gravities by
considering Lie-group valued indices \cite{Boulatov:1992vp,Ooguri:1992eb,DePietri:1999bx}.
These models with group indices, so called group field theory, are actively studied
with interesting recent progress \cite{Carrozza:2011jn}-\cite{Gurau:2009tw}.

The tensor models discussed in this paper are the simplest ones, which have a tensor with 
three indices as their only dynamical variable. By identifying the rank-three tensor as 
the structure constant of the algebra of functions on a fuzzy space,
the rank-three tensor models can be regarded as theory of a dynamical fuzzy space 
\cite{Sasakura:2005js,Sasakura:2011ma}. 
In the previous paper \cite{Sasakura:2011ma}, it has been shown that the rank-three tensor models can 
be described algebraically and 
their symmetry transformations are generated by 3-ary transformations.

3-ary algebras \cite{FigueroaO'Farrill:2008bd,deAzcarraga:2011sh} 
have first been introduced in physics by Nambu \cite{Nambu:1973qe}, and have recently
attracted much attention in the context of M-theory 
\cite{Bagger:2006sk,Gustavsson:2007vu,Bagger:2007jr}.
Higher n-ary algebras have also appeared in the description of 
various dimensional fuzzy spaces \cite{DeBellis:2010pf}.
Since the rank-three tensor models are expected to describe fuzzy spaces
corresponding to any dimensional continuous spaces \cite{Sasakura:2005gv}-\cite{Sasakura:2010rb}, 
it would be natural that higher n-ary algebras also appear in the
rank-three tensor models.
The main purpose of this paper is to point out that, in fact, higher n-ary transformations
also generate the symmetry of the rank-three tensor models.

This paper is organized as follows. In the following section, the correspondence
between the rank-three tensor models and fuzzy spaces is recapitulated. 
The key property is the cyclic condition for the algebra of functions on a fuzzy space,
which corresponds to the generalized hermiticity condition in the rank-three tensor models.
In Section \ref{sec:generalconstruction}, the general method is shown of 
constructing the n-ary algebras which generate the symmetry transformations
of the rank-three tensor models. In Section \ref{sec:violation}, it is discussed 
how the Lie algebraic structure of the symmetry of the rank-three tensor models
is incorporated in the hierarchical structure of the n-ary algebras.
In Section \ref{sec:Leibnitz}, Leibnitz rules are shown to be satisfied by the
n-ary transformations which generate the symmetry of the underlying algebra. 
Then they are shown to form closed finite n-ary Lie subalgebras.
In Section \ref{sec:symmetry}, the roles of the higher n-ary algebras in the tensor models
are discussed. Higher n-ary transformations are shown to 
correspond to more non-local infinitesimal symmetry transformations of the rank-three tensor models.
The final section is devoted to summary and future prospects.

\section{Tensor models and corresponding algebras}
\label{sec:tensormodels}
In this section, I will briefly recapitulate the correspondence between the rank-three tensor models and fuzzy 
spaces, which has been presented in \cite{Sasakura:2011ma}.
 
The tensor models \cite{Ambjorn:1990ge,Sasakura:1990fs,Godfrey:1990dt}
in this paper have a complex dynamical variable 
$M_{abc}\ (a,b,c=1,2,\ldots,N)$, 
which is a tensor with three indices and satisfies the generalized hermiticity condition,
\be
M_{abc}=M_{bca}=M_{cab}=M^*_{bac}=M^*_{acb}=M^*_{cba}, 
\label{eq:ghermiticity}
\ee
where $*$ denotes complex conjugation.
As degrees of freedom, this complex dynamical variable is equivalent to 
the following real dynamical variable defined by
\be
C_{abc}=M_{abc}+M_{bac}+i (M_{abc}-M_{bac}),
\label{eq:cfromm}
\ee
where $C_{abc}$ satisfies the cyclicity condition,
\be
C_{abc}=C_{bca}=C_{cab}.
\label{eq:ccyclic}
\ee

A fuzzy space in this paper is  
characterized by the algebra of real functions $\{ \phi_a | a=1,2,\ldots,N\}$ on it as
\be
\phi_a \phi_b=f_{ab}{}^c\phi_c,
\label{eq:fuzzyalg}
\ee
where $f_{ab}{}^c$ is the real structure constant of the algebra. 
I also assume there exists a metric,
\be
\langle \phi_a|\phi_b\rangle=h_{ab},
\label{eq:metricalg}
\ee
where $<\ |\ >$ is real, symmetric, $h_{ab}=h_{ba}$, and bilinear.

The correspondence between the rank-three tensor models and the fuzzy space is 
assumed to be given by
\be
C_{abc}=f_{ab}{}^{c'}h_{c'c}.
\label{eq:cfrel}
\ee 
Then the cyclicity condition \eq{eq:ccyclic} of $C_{abc}$ 
implies the following cyclicity condition for the algebra and the metric,  
\be
\label{eq:cyclicalg}
\langle \phi_a \phi_b |\phi_c \rangle=\langle \phi_b \phi_c | \phi_a\rangle=\langle \phi_c \phi_a | \phi_b \rangle
=\langle \phi_a| \phi_b \phi_c \rangle
=\langle \phi_b | \phi_c \phi_a \rangle
=\langle \phi_c | \phi_a \phi_b \rangle.
\ee
This is the key property for the following discussions.

\section{General construction of metric invariant n-ary transformations}
\label{sec:generalconstruction}
In the previous paper \cite{Sasakura:2011ma},
the infinitesimal linear transformations which keep the metric \eq{eq:metricalg}
invariant are shown to generate the symmetry of the rank-three tensor models,
and a 3-ary transformation of this kind has been given explicitly.
In this section, I will discuss the extension to n-ary transformations.

For illustration, let me start with reviewing the usual commutator 
from the viewpoint of this paper.
By regarding the indices $b,c$ of $f_{ab}{}^c$ as matrix indices,
the algebra \eq{eq:fuzzyalg} can be regarded as a linear transformation expressed as 
\be
(\phi_a;\phi_b)\equiv \phi_a \phi_b =f_{ab}{}^c \phi_c,
\label{eq:2-1}
\ee
where $\phi_a$ specifies the linear transformation, and $\phi_b$ is the objective
of the linear transformation.
From the property \eq{eq:cyclicalg}, one finds that
\be
\langle \phi_b | (\phi_a;\phi_c) \rangle=\langle \phi_b | \phi_a \phi_c\rangle
=\langle \phi_b \phi_a | \phi_c\rangle=\left\langle \left. \overline{(\phi_a;\phi_b)} \right| \phi_c \right\rangle,
\label{eq:2-2}
\ee
where $\overline{(\ ;\ )}$ denotes the transpose linear transformation given by
\be
\overline{(\phi_a ;\phi_b )}\equiv \phi_b \phi_a.
\label{eq:2-3}
\ee
Here again $\phi_a$ specifies the linear transformation and 
$\phi_b$ is regarded as the objective of the transpose linear transformation.
The transpose of a transpose is identical, because, due to $h_{ab}=h_{ba}$,
the bra and ket states in \eq{eq:2-2} can be converted as 
\be
\langle (\phi_a;\phi_c)|\phi_b \rangle
=\left \langle \phi_c \left| \overline{(\phi_a;\phi_b)}\right. \right\rangle.
\label{eq:2-4}
\ee 
This is expressed as 
\be
\overline{\overline{(\phi_a;\phi_b)}}=(\phi_a;\phi_b).
\ee

Now let me define a linear transformation $[\ ;\ ]$ by 
\be
[\phi_a;\phi_b]\equiv (\phi_a;\phi_b)-\overline{(\phi_a;\phi_b)}.
\label{eq:defof[]}
\ee
Then one can show, from \eq{eq:2-2} and \eq{eq:2-4}, the invariance of the metric,
\be
\langle \phi_b |[\phi_a;\phi_c] \rangle =-\langle [\phi_a;\phi_b]|\phi_c\rangle,
\ee
under the infinitesimal linear transformation $[\ ;\ ]$. 
In fact, this linear transformation 
is the usual commutator $[\phi_a;\phi_b]=\phi_a\phi_b-\phi_b\phi_a$. 

The above procedure can be extended to products of any number of $\phi_a$'s,
\be
(\phi_{a_1},\phi_{a_2},\ldots,\phi_{a_{p}},s;\phi_b)=\hbox{product of }\phi_{a_1},\phi_{a_2},
\ldots,\phi_{a_{p}},\phi_b\hbox{ in way $s$}.
\label{eq:naryprod}
\ee
Here $\phi_{a_1},\phi_{a_2},\ldots,\phi_{a_{p}}$ specify the linear transformation,
$\phi_b$ is regarded as the objective of the linear transformation, 
and $s$ symbolically specifies the way of product. 
By moving $\phi_{a_i}$'s from the ket state to the bra state
with the use of the cyclic property \eq{eq:cyclicalg}, one can 
obtain the transpose linear transformation satisfying 
\be
\left\langle \phi_b \left| (\phi_{a_1},\phi_{a_2},\ldots,\phi_{a_{p}},s;\phi_c)\right. \right\rangle
=\left\langle \left.(\phi_{a_1},\phi_{a_2},\ldots,\phi_{a_{p}},\overline{s};\phi_b)\right|\phi_c 
\right\rangle, 
\label{eq:deftranspose}
\ee 
where $\overline{s}$ symbolically denotes the transpose, namely,
\be
(\phi_{a_1},\phi_{a_2},\ldots,\phi_{a_{p}},\overline{s};\phi_b)
\equiv\overline{(\phi_{a_1},\phi_{a_2},\ldots,\phi_{a_{p}},s;\phi_b)}.
\ee
As above, the transpose of a transpose is identical, 
\be
\overline{\overline{s}}=s,
\label{eq:transposeoftranspose}
\ee
which is a direct consequence of $h_{ab}=h_{ba}$.

Now let me define 
\be
[\phi_{a_1},\phi_{a_2},\ldots,\phi_{a_{p}},s;\phi_b]=
(\phi_{a_1},\phi_{a_2},\ldots,\phi_{a_{p}},s;\phi_b)-
(\phi_{a_1},\phi_{a_2},\ldots,\phi_{a_{p}},\overline{s};\phi_b).
\label{eq:defof[n]}
\ee
From \eq{eq:deftranspose} and \eq{eq:transposeoftranspose}, one can show that the
metric \eq{eq:metricalg} is invariant under the linear transformation \eq{eq:defof[n]} as
\be
\left\langle \left. [\phi_{a_1},\phi_{a_2},\ldots,\phi_{a_{p}},s;\phi_b]\right| \phi_c\right\rangle
=-\left\langle  \phi_b \left|[\phi_{a_1},\phi_{a_2},\ldots,\phi_{a_{p}},s;\phi_c] \right. \right\rangle.
\label{eq:invmetn}
\ee
From \eq{eq:transposeoftranspose} and \eq{eq:defof[n]}, it is obvious that
\be
 [\phi_{a_1},\phi_{a_2},\ldots,\phi_{a_{p}},s;\phi_b]=-
[\phi_{a_1},\phi_{a_2},\ldots,\phi_{a_{p}},\overline{s};\phi_b].
\ee

The 3-ary transformation
 studied in the previous paper \cite{Sasakura:2011ma} is just a special case of the above general
procedure. In the case, the starting linear transformation is taken to be 
\be
(\phi_a,\phi_b;\phi_c)=(\phi_a\phi_c)\phi_b.
\ee
By applying the above procedure, one obtains
\bea
\overline{(\phi_a,\phi_b;\phi_c)}&=&(\phi_b\phi_c)\phi_a, \nonumber \\
{[} \phi_a,\phi_b;\phi_c ] &=&(\phi_a,\phi_b;\phi_c)-\overline{(\phi_a,\phi_b;\phi_c)}
=(\phi_a\phi_c)\phi_b-(\phi_b\phi_c)\phi_a,
\label{eq:3ary[]}
\eea
which is indeed the 3-ary transformation studied in \cite{Sasakura:2011ma}. 

\section{Violation of Leibnitz rules and hierarchy of n-ary algebras}
\label{sec:violation}
Since the infinitesimal linear transformations which keep the metric $h_{ab}$ invariant form a Lie algebra,
it would be interesting to discuss the Lie algebraic structure of the above n-ary transformations.
Let me start with the following product of $(p+1)$- and $(q+1)$-ary linear transformations,
\be
(\phi_{a_1},\phi_{a_2},\ldots,\phi_{a_{p+q}},s*t;\phi_b)\equiv
[\phi_{a_1},\phi_{a_2},\ldots,\phi_{a_p},s;[\phi_{a_{p+1}},\phi_{a_{p+2}},\ldots,\phi_{a_{p+q}},t;\phi_b]].
\label{eq:prodnm}
\ee
In general, this product is not an infinitesimal transformation which keeps the metric $h_{ab}$ invariant. 
The transpose of the transformation is given by
\be
(\phi_{a_1},\phi_{a_2},\ldots,\phi_{a_{p+q}},\overline{s*t};\phi_b)=
[\phi_{a_{p+1}},\phi_{a_{p+2}},\ldots,\phi_{a_{p+q}},t;[\phi_{a_1},\phi_{a_2},\ldots,\phi_{a_p},s;\phi_b]].
\ee
This is because
\bea
\left\langle \phi_b \left| (\phi_{a_1},\phi_{a_2},\ldots,\phi_{a_{p+q}},s*t;\phi_c)\right.\right\rangle
&=&\left\langle \phi_b \left| 
[\phi_{a_1},\phi_{a_2},\ldots,\phi_{a_p},s;[\phi_{a_{p+1}},\phi_{a_{p+2}},\ldots,\phi_{a_{p+q}},t;\phi_c]]
\right.\right\rangle \nonumber \\
&=&
-\left\langle [\phi_{a_1},\phi_{a_2},\ldots,\phi_{a_p},s;\phi_b] \left| 
[\phi_{a_{p+1}},\phi_{a_{p+2}},\ldots,\phi_{a_{p+q}},t;\phi_c]
\right.\right\rangle \nonumber \\
&=&
\left\langle \left. [\phi_{a_{p+1}},\phi_{a_{p+2}},\ldots,\phi_{a_{p+q}},t;
[\phi_{a_1},\phi_{a_2},\ldots,\phi_{a_p},s;\phi_b]] \right| \phi_c
\right\rangle, \nonumber \\ 
\eea
where I have used \eq{eq:invmetn}.
Therefore the commutator of the linear transformations, $s$ and $t$, has the form discussed in the previous section,
\bea
&&[\phi_{a_1},\ldots,\phi_{a_p},s;[\phi_{a_{p+1}},\ldots,\phi_{a_{p+q}},t;\phi_b]]
-[\phi_{a_{p+1}},\ldots,\phi_{a_{p+q}},t;[\phi_{a_1},\ldots,\phi_{a_p},s;\phi_b]]
\nonumber \\
&&\ \ \ \ \ \ \ \ \ \ \ \ \ \ \ \ \ =\ (\phi_{a_1},\phi_{a_2},\ldots,\phi_{a_{p+q}},s*t;\phi_b)-
(\phi_{a_1},\phi_{a_2},\ldots,\phi_{a_{p+q}},\overline{s*t};\phi_b).
\label{eq:comofst}
\eea
This shows that the commutator gives an infinitesimal linear transformation which keeps the metric invariant. 
This is of course expected, 
because the infinitesimal linear transformations which keep $h_{ab}$ invariant form a Lie algebra. 

The exact form of the right-hand side of \eq{eq:comofst} depends on the  structure of the algebra \eq{eq:fuzzyalg}. 
If the n-ary transformations, $s$ and $t$, satisfy the Leibnitz rule (or so called fundamental identity) given by
\bea
[\phi_{a_1},\ldots,\phi_{a_p},s;[\phi_{a_{p+1}},\ldots,\phi_{a_{p+q}},t;\phi_b]]
&=&[[\phi_{a_1},\ldots,\phi_{a_p},s;\phi_{a_{p+1}}],\phi_{a_{p+2}},\ldots,\phi_{a_{p+q}},t;\phi_b]
\nonumber \\
&&+[\phi_{a_{p+1}}, [\phi_{a_1},\ldots,\phi_{a_p},s;\phi_{a_{p+2}}],\ldots,\phi_{a_{p+q}},t;\phi_b]
\nonumber \\
&&+ \cdots
\nonumber \\
&&+[\phi_{a_{p+1}},\ldots,\phi_{a_{p+q}},t; [\phi_{a_1},\ldots,\phi_{a_p},s;\phi_b]], 
\eea
then the right-hand side of \eq{eq:comofst} can be expressed by $s$ or $t$ linear transformations, and 
the n-ary transformations will close by themselves.
In general, however, the Leibnitz rules do not hold, and 
the right-hand side of \eq{eq:comofst} will be given by a linear summation of $(p+q+1)$-ary transformations,
\be
\sum_{s} c_s \, [\phi_{a_1},\cdots,\phi_{a_{p+q}},s;\phi_b],
\ee
where each $[\ ;\ ]$ has the form defined in \eq{eq:defof[n]}, and $c_s$ are real coefficients.
Thus to incorporate the Lie algebraic structure of the metric invariant n-ary transformations, it is in general 
necessary to consider the whole hierarchy of the n-ary transformations of the form \eq{eq:defof[n]}.

\section{Condition for Leibnitz rules}
\label{sec:Leibnitz}
In this section, I will discuss the condition for the Leibnitz rules to hold from the perspective
of the invariance of the underlying algebra.

The $(p+1)$-ary linear transformation defined in \eq{eq:defof[n]} may be expressed as
a tensorial form like
\be
[\phi_{a_1},\phi_{a_2},\ldots,\phi_{a_{p}},s;\phi_b]=M^s_{a_1\ldots a_{p}\,b}{}^c\phi_c,
\label{eq:narym}
\ee 
where $M^s_{a_1\ldots a_{p}\,b}{}^c$ are real.
With $M^s_{a_1\ldots a_{p}\,b}{}^c$, 
the invariance of the metric $h_{ab}$ under the infinitesimal linear transformation \eq{eq:narym}
can be expressed as
\bea
0&=&
\left\langle \left. [\phi_{a_1},\phi_{a_2},\ldots,\phi_{a_{p}},s;\phi_b]\right|
\phi_c \right\rangle+
 \left\langle \phi_b \left |  [\phi_{a_1},\phi_{a_2},\ldots,\phi_{a_{p}},s;\phi_c]\right.
\right\rangle
\nonumber \\
&=&
M^s_{a_1\ldots a_{p}\,b}{}^{b'} h_{b'c}+M^s_{a_1\ldots a_{p}\,c}{}^{c'} h_{bc'}.
\eea
This implies the antisymmetry of the last two indices of $M$,
\be
M^s_{a_1\ldots a_{p}\,bc}=-M^s_{a_1\ldots a_{p}\,cb},
\ee
where $M^s_{a_1\ldots a_{p}\,bc}\equiv M^s_{a_1\ldots a_{p}\,b}{}^d h_{dc}$.

Now let me assume that the algebra \eq{eq:fuzzyalg} is invariant under the infinitesimal 
linear transformation \eq{eq:narym} for a specific choice of $\phi_{a_1},\phi_{a_2},\ldots,\phi_{a_{p}},s$.
It is straightforward to extend the following discussions for linear combinations of $(p+1)$-ary linear 
transformations which keep \eq{eq:fuzzyalg} invariant.
The assumption can be expressed by the following invariance of the structure constant of the algebra,
\be
M^s_{a_1\ldots a_{p}\,b}{}^{b'} f_{b'c}{}^d+M^s_{a_1\ldots a_{p}\,c}{}^{c'} f_{bc'}{}^d
-M^s_{a_1\ldots a_{p}\,d'}{}^{d} f_{bc}{}^{d'}=0.
\label{eq:leibnitz1}
\ee
Or, in an algebraic form, this is equivalent to the Leibnitz rule,
\be
[\phi_{a_1},\phi_{a_2},\ldots,\phi_{a_{p}},s;\phi_b \phi_c]
=[\phi_{a_1},\phi_{a_2},\ldots,\phi_{a_{p}},s;\phi_b]\phi_c+
\phi_b [\phi_{a_1},\phi_{a_2},\ldots,\phi_{a_{p}},s;\phi_c]
\label{eq:leibnitz2}
\ee
for any $\phi_b$ and $\phi_c$.

By iteratively applying \eq{eq:leibnitz2}, 
it is obvious that the objective of the linear transformation can be any product of any number of $\phi$'s
for the Leibnitz rules to hold.
Therefore, the Leibnitz rules for the n-ary transformations, or so called fundamental identities, hold as
\bea
[\phi_{a_1},\ldots,\phi_{a_{p}},s;[\phi_{b_1},\ldots,\phi_{b_q},t;\phi_c]]
&=&[[\phi_{a_1},\ldots,\phi_{a_{p}},s;\phi_{b_1}],\ldots,\phi_{b_q},t;\phi_c]
\nonumber \\
&&+[\phi_{b_1}, [\phi_{a_1},\ldots,\phi_{a_{p}},s;\phi_{b_2}],\ldots,\phi_{b_q},t;\phi_c]
\nonumber \\
&&+ \cdots
\nonumber \\
&&+[\phi_{b_1},\ldots,\phi_{b_q},t; [\phi_{a_1},\ldots,\phi_{a_{p}},s;\phi_c]].
\eea 
Here it is important to note that $\phi_{b_1},\ldots,\phi_{b_q},t,\phi_c$ can be arbitrary,
while $\phi_{a_1},\ldots,\phi_{a_{p}},s$ must be taken so that the linear 
transformation \eq{eq:narym} be the symmetry of the algebra and satisfy \eq{eq:leibnitz1} or \eq{eq:leibnitz2}.

Interesting structures appear when both 
$[\phi_{a_1},\phi_{a_2},\ldots,\phi_{a_{p}},s;\ ]$ and 
$[\phi_{b_1},\ldots,\phi_{b_q},t;\ ]$ are taken to be the infinitesimal 
linear transformations being the symmetry of the underlying algebra.
Then, as discussed in the previous section, due to the Leibnitz rule,
the commutator of these linear transformations
will not generate other n-ary linear transformations.
Since the infinitesimal linear transformations which generate the symmetry of the underlying algebra
form a Lie algebra, the n-ary linear transformation obtained from the commutator is again
a symmetry transformation.
This implies that the n-ary transformations which generate the symmetry of the underlying algebra
will form closed finite n-ary Lie subalgebras.

\section{The rank-three tensor models and n-ary algebras}
\label{sec:symmetry}
The symmetry of the rank-three tensor models is the orthogonal group symmetry,
\be
C_{abc}\rightarrow O_{a}{}^{a'} O_{b}{}^{b'}O_{c}{}^{c'}C_{a'b'c'},\ \ \ O\in O(N,R).
\label{eq:otransc}
\ee
As discussed in \cite{Sasakura:2011ma}, 
this orthogonal group symmetry of the tensor models can be identified with 
the symmetry of fuzzy spaces by partially gauge-fixing the latter symmetry 
with the gauge fixing condition, 
\be
h_{ab}=\delta_{ab}.
\label{eq:hdelta}
\ee
Then indeed the n-ary linear transformations \eq{eq:defof[n]} become the infinitesimal 
orthogonal group transformations,
which keep $h_{ab}=\delta_{ab}$ invariant, and can be identified with the generators of the symmetry
transformations \eq{eq:otransc} of the rank-three tensor models. 

In the previous paper \cite{Sasakura:2011ma}, it is argued that the 3-ary transformations,
\be
\delta_{bc} \phi_a = [\phi_b,\phi_c;\phi_a]
\label{eq:deltabc}
\ee
with $[\ ; \ ]$ defined in \eq{eq:3ary[]},
will generate the symmetry transformations of the tensor models
by taking various choices of the functions $\phi_b$ and $\phi_c$.
Since the number of the independent elements of the Lie algebra
$so(N)$ agrees with the number of independent choices of $\phi_b$ and $\phi_c$ on account
of the antisymmetry of the first two entries of the 3-ary transformation,
the 3-ary transformations \eq{eq:deltabc} will generate all the infinitesimal symmetry transformations of the tensor 
models, unless $C_{abc}$ is fine-tuned not to be so.

The above statement is mathematically true, but this is not physically realistic. 
To explain this, for illustration, let me consider a class of $C_{abc}$ with the following Gaussian forms,
\be
C_{x_1 x_2 x_3}=\exp \left[
-\beta\left((x_1-x_2)^2+(x_2-x_3)^2+(x_3-x_1)^2\right)\right],
\label{eq:cgauss}
\ee
where $\beta$ is a positive real parameter with dimension (length)$^{-2}$, the indices $x_i$'s are
$D$-dimensional coordinates, and 
\be
x^2\equiv g_{\mu\nu} x^\mu x^\nu,
\ee 
with a real symmetric two-tensor $g_{\mu\nu}$. The $g_{\mu\nu}$ is assumed to be a positive definite matrix. 
Physically the algebra\footnote{The repeated index is assumed to be integrated over.}
\be
\phi_{x_1} \phi_{x_2}=C_{x_1 x_2 x_3} \phi_{x_3},
\ee
which is obtained from \eq{eq:cgauss}
through \eq{eq:cfrel} and \eq{eq:fuzzyalg} with \eq{eq:hdelta},
can be regarded as the algebra of the functions on 
a $D$-dimensional fuzzy flat space with fuzzy scale $1/\sqrt{\beta}$ \cite{Sasai:2006ua}.
Here the function $\phi_x$ may be regarded as a fuzzy analogue of the delta function $\delta^D(z-x)$ on a usual 
continuous space.

With $\phi_x$, the 3-ary transformations are given by
\be
\delta_{x_2x_3} \phi_{x_1} = [\phi_{x_2},\phi_{x_3};\phi_{x_1}].
\label{eq:3aryx}
\ee
Indeed one can numerically\footnote{The continuous $x_i$'s are discretized
for the numerical check, and only the case $D=1$ is checked.} check 
that the 3-ary transformations \eq{eq:3aryx} generate all the infinitesimal symmetry 
transformations of the tensor models by taking all the choices of $\phi_{x_2}$ and $\phi_{x_3}$.
However, it is obvious, from the Gaussian damping form of \eq{eq:cgauss}, that, 
for the 3-ary transformations \eq{eq:3aryx} to be significant,
$x_2$ and $x_3$ must be taken within the range of order $1/\sqrt{\beta}$ around $x_1$.
Thus, practically,  the 3-ary transformations \eq{eq:3aryx} only generate
local (with fuzziness $1/\sqrt{\beta}$) infinitesimal transformations, while the Lie generators for \eq{eq:otransc}
of the tensor models
contain non-local ones $o_{x_1}{}^{x_2}\in so(N)$ with arbitrary choices of $x_1$ and $x_2$.

The above conclusion does not depend on the choice of a basis of functions on a fuzzy space, 
nor is it restricted to the specific Gaussian form \eq{eq:cgauss}. 
Any physically acceptable fuzzy space must respect locality (with certain fuzziness), and a basis of functions may
be taken so that each function be localized within a characteristic scale of fuzziness. 
Then products of two functions become significant only when the two functions are 
overlapping within the characteristic scale.
Thus, practically, for general fuzzy spaces respecting locality, 
the 3-ary transformations can only generate local infinitesimal transformations, which are far less than
all the infinitesimal symmetry transformations of the tensor models.

The other more non-local infinitesimal symmetry transformations will be obtained by considering higher  
n-ary transformations defined in \eq{eq:defof[n]}. A $p$-ary transformation contains
$(p-1)$ products of functions, and can be made significant by ordering the localized basis functions 
in a chain so that neighboring functions are overlapping within the characteristic scale of fuzziness. 
Since such a chain can be made arbitrarily long by considering an arbitrary number of functions, 
arbitrarily non-local infinitesimal symmetry transformations will be obtained 
by considering arbitrarily higher n-ary transformations.    
The necessity of higher n-ary transformations is also natural
from the Lie algebraic structure of the symmetry of the tensor models,
 as discussed in Section \ref{sec:violation},
 because successive applications of the local infinitesimal transformations will generate
 non-local transformations.

\section{Summary and future prospects}
\label{sec:summary}
In this paper, I have discussed the n-ary algebras for describing the symmetry of the tensor models.
I have presented the general method to construct the n-ary transformations which generate the 
symmetry of the tensor models. The 3-ary algebra reported in the previous paper is
just a particular case of the general method.
The Lie algebraic structure of the symmetry of the tensor models
can be incorporated by the hierarchical structure of the n-ary algebras.
If there exist n-ary transformations which generate the symmetry of the underlying algebra, 
the Leibnitz rules hold for
them, and they form closed finite n-ary Lie subalgebras.
The 3-ary transformations can generally generate all the infinitesimal
symmetry transformations of the tensor models, but
in physical settings, they practically generate only local infinitesimal symmetry transformations
due to the locality which should be respected by physically acceptable fuzzy spaces. 
The other more non-local
infinitesimal symmetry transformations of the tensor models can be generated by higher n-ary transformations.

The starting key assumption of this paper is the cyclic property concerning the algebra of functions on a fuzzy
space and its metric. This property allows systematic construction of the symmetry generators.
The property has been derived in the context of the tensor models, but it is also interesting 
to set this assumption in the general context of fuzzy spaces.
The notion of fuzzy space is an interesting candidate to replace 
the classical spacetime notion of general relativity.
However, this new notion is currently lacking a general basic framework and is 
yet to be developed.
The settings assumed for fuzzy spaces in this paper may hopefully shed new light in this direction.


\end{document}